\documentstyle[prl,aps,multicol,epsf]{revtex}
\begin{document} 
\title{Observation of Quantum Fluctuations of Charge on a Quantum Dot}
\author{D. Berman, N. B. Zhitenev and R. C. Ashoori}
\address{Department of Physics, Massachusetts Institute of
Technology, Cambridge, Massachusetts, 02139}
\author{M. Shayegan}
\address{ Department of Electrical Engineering, Princeton University, 
Princeton NJ, 08544}
\maketitle
\date{Received}
\begin{abstract}
We have incorporated an aluminum single electron transistor
directly into the defining gate structure of a semiconductor quantum dot,
permitting precise measurement of the charge in the dot. Voltage biasing
a gate draws charge from a reservoir into the dot through
a single point contact. The charge in the dot increases continuously for
large point contact conductance and in a step-like manner in units of single 
electrons with the contact nearly closed.  We measure the corresponding 
capacitance lineshapes for the full range of point contact conductances. The
lineshapes are described well by perturbation theory and not by theories 
in which the dot charging energy is altered by the barrier conductance.
\end{abstract}
\begin{multicols}{2}
\narrowtext
In classical physics, a puddle of electrons holds a discrete and 
measurable number of electrons. Quantum mechanics instead dictates 
that the probability for an electron to be in a localized state on 
the puddle depends on the coupling strength to the environment. 
For many systems in which a single state is coupled to a continuum, 
this coupling produces a ``lifetime broadening" of energy levels. 
For instance, atomic spectra display a characteristic Lorentzian 
lineshape broadening \cite{Weisskopf}. In analogy with atomic 
spectroscopy, several experiments have demonstrated the capability 
of precisely measuring the energies to add electrons to quantum dots 
\cite{overview}. In contrast to atomic physics, the lineshape of 
quantum dot levels originates essentially in a many-body interaction 
between electrons in the dot and the macroscopic environment.

As the tunnel barrier conductance, $G$, between the quantum dot and 
the macroscopic leads is increased above $2e^{2}/h$, charge is no 
longer quantized and the Coulomb blockade is destroyed.  This process
has been attributed to quantum charge fluctuations between the dot 
and the environment \cite{Devoret}.  A thorough physical 
description of this effect has only been recently proposed in the 
nearly closed regime $(G \ll e^{2}/h)$ \cite{GlazmanMatveev,weak} 
and the nearly open regime $(2e^{2}/h - G \ll e^{2}/h)$ \cite{strong}. 

Experiments measuring the charge or the capacitance of a dot provide 
the most direct information about charge fluctuations and the effect 
of the dot-environment interaction on the charging states of the dot. 
However, transport measurements have been the first to address the 
issue of dot-environment coupling. In one of the first studies, Foxman 
et. al. \cite{Foxman} examined the lineshape of conductance peaks with 
increasing coupling of the dot to the leads and found good agreement 
with Lorentzian broadening. To analyze the charging lineshapes in the 
dot for a broad range of coupling strengths, conductance measurements 
are poorly suited, being complicated by other processes such as cotunneling 
\cite{Furusaki} and Kondo coupling \cite{Goldhaber}.

Previous experiments have addressed the issue of charging lineshapes. 
Experimenters employed a semiconductor electrometer \cite{Molenkamp} 
to observe the effect of charge fluctuations. They modeled their results by 
a reduction of the charging energy with increasing coupling.  In another 
experiment, the effect of tunnel  barrier conductance on Coulomb 
blockade was studied through peak splitting of double dots \cite{Livermore}.  
In this case, the spacing between double-dot peaks can be predicted with a 
similar formalism as we use in our lineshape analysis \cite{MatveevGlazman}.

We have developed an experiment which probes the capacitance lineshape 
of a quantum dot with unprecedented sensitivity. We find that the lineshapes 
deviate substantially from previously employed fitting forms 
\cite{Foxman,Molenkamp} and are best described for all coupling strengths 
by the theory developed recently by Matveev \cite{weak,strong}.

We measure the capacitance lineshapes of a quantum dot with only one 
contact to a charge reservoir.  The quantum dot is electrostatically 
defined in a two-dimensional electron gas (2DEG) of a AlGaAs/GaAs 
heterostructure. The 2DEG is about 1200$\AA$ below the surface with a carrier 
concentration of $1 \times 10^{11}$cm$^{-2}$. Measurements were performed on 
six different samples, each yielding very similar results, and here we present 
detailed data from one of them. A micrograph of the structure is shown in Fig. 
1a. The estimated area of the quantum dot is about 0.5$\mu m^{2}$, which 
corresponds to an energy level spacing of 7$\mu$eV.  We measured the average 
charging energy of the dot to be $U=e^{2}/2C_{\Sigma}=0.23$meV from temperature 
dependence of the capacitance peaks for high tunneling barriers. Here, 
$C_{\Sigma}$=348aF is the total capacitance of the quantum dot.  These
parameters agree with what we expect from geometrical considerations.

The charge on the quantum dot is measured with a single-electron transistor 
(SET) with extremely high sensitivity \cite{Likharev}. The metal SET is 
fabricated \cite{Berman} with Al-Al$_{2}$O$_{3}$-Al tunnel junctions 
using the standard shadow-evaporation method \cite{Fulton}. To maximize the 
sensitivity to the quantum dot charge, we incorporate the SET directly into one 
of the leads defining the dot. 

Figure 1b. shows the drain-source current-voltage relationship of the SET.
It changes cyclically with the charge induced on the central island of the SET. 
The dependence of the current on the SET central island charge is shown in 
Fig. 1c. For optimal charge sensitivity of the SET, we set the drain-source 
voltage at the onset of conduction for the maximum Coulomb blockade 
condition \cite{Lafarge}, as shown by the arrow in Fig. 1b.  For the sample 
primarily discussed in this Letter, we achieve a sensitivity of 
$1.2 \times 10^{-3} e/ \sqrt{Hz}$ to the quantum dot charge.
\begin{figure}
\epsfxsize=\linewidth
\epsfbox{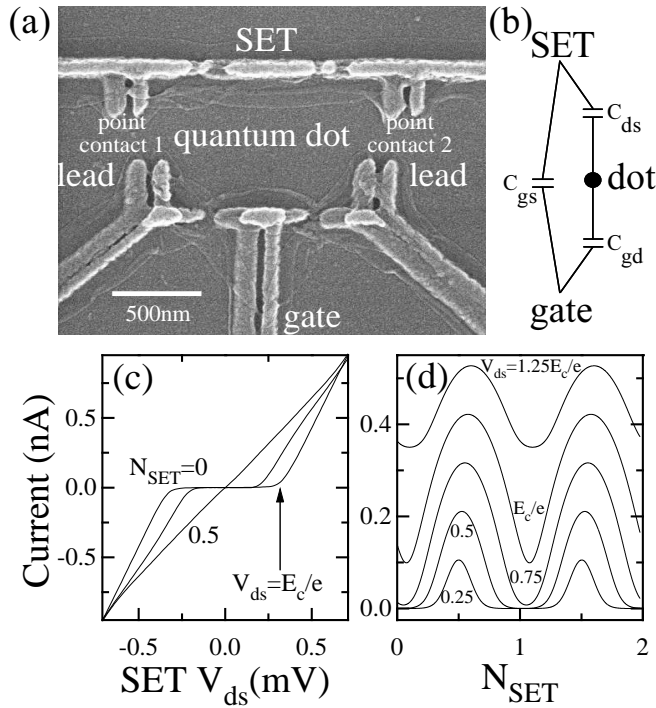}
\caption{(a) Micrograph of measurement setup.  The leads are made of aluminum 
by shadow evaporation.  The area of the quantum dot is approximately 
0.5$\mu m^{2}$. (b) Schematic of some of the capacitances in the measurement.
(c) Example of drain-source current-voltage characteristics of a single-electron 
transistor at a refrigerator temperature of 50mK shown for three values of gate 
voltage $C_{gs}V_{g}=eN_{SET}$=0, 0.25 and 0.5 electrons. The arrow shows the 
drain voltage bias for optimal gain. (d) Dependence of the SET current with 
transparent quantum dot tunnel barriers on gate voltage for different 
drain-source voltage biases. Maximum peak-to-valley modulation amplitude is
at SET $V_{ds}=E_{c}/e$.}
\end{figure}
Through application of a DC voltage, $V_g$, to the lead marked ``gate" in Fig. 1a, 
charge can be drawn onto the dot as $eN=C_{gd}V_g$, where $ C_{gd}$ is
the gate-dot capacitance.  However, for zero temperature and for high tunneling 
barriers separating the dot from the leads, the charge on the quantum dot is
quantized and can only change from $n$ to $n+1$ around points in gate voltage, 
where $N=(n+0.5)$. The measured capacitance is 
$C_{meas}=e \partial \overline{n} / \partial V_g$, where $\overline{n}$ is the average 
number of electrons on the dot.  

The capacitance lineshape is measured by applying a small ac excitation 
(40 $\mu$V {\it rms}, 1kHz) to the gate. This signal modulates the charge on 
the quantum dot by an amount that is a function of $N$ and the coupling 
strength.  The small ac modulation of the quantum dot charge induces ac 
charge on the SET central island resulting in a current through the 
SET at the excitation frequency. Examples of measured SET response as 
$V_g$ is swept are shown in Fig. 2a for three different tunnel coupling 
strengths.  The upper trace is obtained for $G=1.65e^{2}/h$, where 
$\overline{n}$ deviates only slightly from $N$ and the electrostatic potentials 
in the dot and the leads are nearly equal. A prominent feature of this curve is an 
oscillation with a period of 94mV.  This period arises due to an 
addition of one electron to the SET central island through a direct 
capacitance $C_{gs}=1.7$aF to the gate, modulating the gain of the SET. 
\begin{figure}
\epsfxsize=\linewidth
\epsfbox{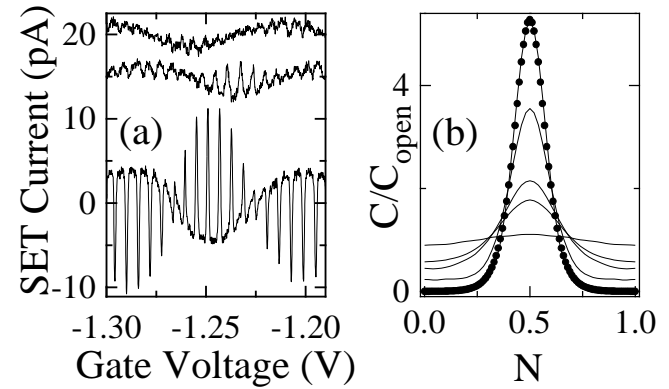}
\caption{(a) SET signal vs gate voltage for three values of point contact 
conductance.  Top to bottom:  $G$=1.65, 1.32 and 0.05$e^{2}/h$.  (b) Solid lines: 
five capacitance peaks extracted from data: $G$ is varied from 0.010 to 
1.81$e^{2}/h$. Closed circles: derivative of the Fermi function for a temperature 
of 130mK.}
\end{figure}
The bottom trace in Fig. 2a is obtained for $G=0.05 e^{2}/h$.  Here, the 
charge on the dot is well quantized and can only change in close 
proximity to points where $N=(n+0.5)$. These points correspond to the 
sharp peaks in the trace, spaced with a mean period of 6.3mV, yielding a 
gate-dot capacitance of $C_{gd}=25aF$. Notice that the large-period 
background oscillation has a larger amplitude compared with the upper 
traces in Fig. 2a. Between the peaks, the dot potential is effectively 
floating; charge cannot enter the dot from the reservoir to screen the 
ac gate potential. Thus, more charge is induced on the SET in response to 
the ac excitation on the gate because the ac coupling from the gate 
to the SET is augmented by a factor of $C_{gd}C_{ds}/C_{\Sigma}$.  
Here $C_{ds}$ is the quantum dot-SET central island capacitance. 

In general, the charge response on the SET central island, $dQ_{SET}$, to
the ac excitation on the gate, $dV_g$, can be expressed as:
\begin{equation}
dQ_{SET}= \left[ \left( C_{gd}-C_{meas} \right) \frac{C_{ds}}{C_{\Sigma}}+C_{gs} \right]dV_{g}.
\label{eq:SET}
\end{equation}
As our SET operates in the linear response regime, the current through the SET 
directly reflects $dQ_{SET}$. Linear response is ensured because the ratio of 
$C_{ds}$ to the total capacitance of the SET central island is about 0.05. 
Therefore, a change of charge of one electron in the quantum dot only induces 
${1/20}^{\rm th}$ of an electron on the SET. Moreover, we obtain our 
capacitance lineshapes at maximal gains of SET where this small induced charge 
has minimal effect on the SET gain.  The reverse effect of the SET on the 
quantum dot charge is also very small. The ratio $C_{ds}/C_{\Sigma}$ is 
approximately 0.06, producing negligible feedback. Also, the charge on the SET 
central island is poorly quantized since a finite source-drain voltage is applied to 
the SET.  Using equation (\ref{eq:SET}), we extract the quantum dot capacitance 
lineshapes, $C_{meas}(V_{g})$, from the raw data as a function of the tunnel barrier 
conductance.  

During the measurement of the capacitance lineshapes, point contact 2 
is completely pinched off, and the dot is coupled to the leads only through point 
contact 1.  To determine the conductance of contact 1 in this regime, 
we perform the following procedure.  The conductance of contact 1 is 
measured with 2 completely open. To account for the electrostatic 
coupling between contacts 1 and 2,  we monitor the shift of 
conductance plateaus of contact 1 as 2 is being closed.  This 
procedure allows us to extrapolate $G$, the conductance of contact 1, 
to the regime of the capacitance measurement.

Figure 2b shows the evolution of the capacitance lineshape with 
increased coupling strength.  The nominal values of $G$ are:  0.010, 
0.67, 1.09, 1.50 and $1.81e^{2}/h$. It is clear that as $G$ increases 
and approaches $2e^{2}/h$, the capacitance peaks broaden and the Coulomb 
blockade oscillations diminish and disappear. Below, we discuss the 
lineshapes of different quantum dot capacitance peaks in various coupling 
regimes: very weak, weak and strong.

In the very weak coupling regime, the shape of the capacitance peak is 
determined simply by thermal broadening.  Figure 2b shows good agreement 
between a peak measured with $G=0.010e^{2}/h$ and a derivative of the 
Fermi-Dirac function for a temperature of 130mK.

For larger tunnel barrier conductance, the capacitance lineshape 
changes. In Figs. 3a, b, c and d, we plot with open circles 
capacitance peaks that we obtained for nominal values of $G$=0.67, 
1.09, 1.50 and 1.81$e^{2}/h$. We compared our capacitance peaks with 
expressions that have been previously used to fit conductance peaks.  For 
example, Lorentzian lifetime broadening has been considered \cite{Foxman} 
for characterizing the charge smearing effects. In Fig. 3a, the lower 
plot of Fig. 3b, Figs. 3c and d, we plot Lorentzian-broadened 
Fermi peaks with energy level widths $\Gamma$=0.15, 0.32, 0.44 and 
1.0$U$.  The lineshapes show significant deviations from the data.  To avoid 
clutter, we have fit the Lorentzians to the valleys between our peaks.  
Nonetheless, fitting to the peak centers gives an equally poor result.
\begin{figure}
\epsfxsize=\linewidth
\epsfbox{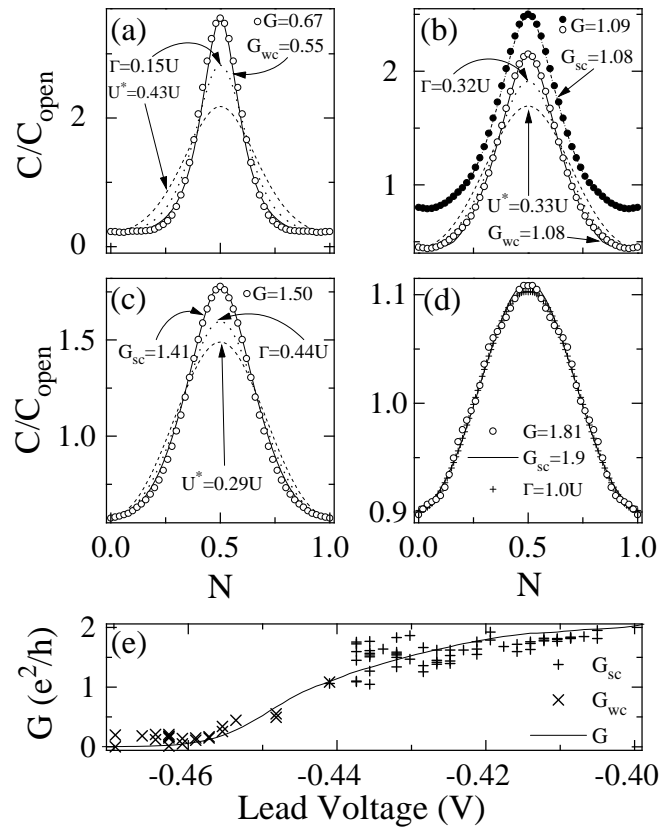}
\caption{(a) Open circles: data for $G=0.67 e^{2}/h$, solid line: fit to weak 
coupling theory (wc) with $G_{wc}=0.55 e^{2}/h$, dotted line: Lorentzian 
with $\Gamma=0.15U$, dashed line: derivative of the Fermi function with 
$U^{*}=0.43U$.  (b) Open circles: data for $G=1.09 e^{2}/h$, solid line: 
wc fit with $G_{wc}=1.08 e^{2}/h$, dotted line: Lorentzian with 
$\Gamma=0.32U$, dashed line: Fermi function with $U^{*}=0.33U$.  Closed circles:  
data for $G=1.09 e^{2}/h$ offset by 0.35 vertically, dash-dot line: strong coupling theory 
(sc) fit with $G_{sc}=1.08 e^{2}/h$.  (c) Open circles: data for $G=1.50 e^{2}/h$, solid line: sc fit with
$G_{sc}=1.41 e^{2}/h$, dotted line: Lorentzian with $\Gamma=0.44U$, dashed line: Fermi 
function with $U^{*}=0.29U$.  (d) Open circles: data for $G=1.81 e^{2}/h$, solid line: sc fit with
$G_{sc}=1.90 e^{2}/h$, crosses: Lorentzian with $\Gamma=1.0U$.  (e) Tunnel barrier 
conductance (solid line) vs tunnel barrier lead voltage.  $\times$:  Conductance values 
obtained from fits with weak coupling theory (wc).  +:  Conductance values obtained from 
fits with strong coupling theory (sc).}
\end{figure}
Previous measurements of charge fluctuations used a renormalized 
charging energy $U^*$ to account for peaks broadened with a finite 
tunnel barrier conductance \cite{Molenkamp}. In Figs. 3a, b and c,
we plot derivatives of the Fermi function with $U^*$=0.43, 0.33 and 
0.29$U$ for a temperature of 130mK.  These lineshapes clearly do not fit the 
data either.

Finally, we compared our experimental results to the theoretical treatment 
developed by Matveev \cite{weak,strong}. The problem of interaction between 
the dot and the leads was solved in the limits of weak \cite{GlazmanMatveev,weak} 
and strong \cite{strong} coupling using either transmission or reflection of the 
tunnel barrier as a small parameter in perturbation theory. In both limits, the 
physics of charge fluctuations is related to spin fluctuations in the Kondo 
problem. Here, instead of the degeneracy of the two-spin states, there is a 
degeneracy between the dot states with $n$ and $n+1$ electrons. Similarly to the 
Kondo effect, the charge displays a logarithmic divergence around these degeneracy 
points at very low temperatures. As a result, the predicted capacitance lineshape 
has more weight around the half integer values of $N$ in comparison with other 
theoretical treatments.

For weak coupling, this effect becomes pronounced at experimentally 
unattainable temperatures.  Therefore, in the range of $G_{wc} \ll e^{2}/h$, 
it suffices to treat the tunnel coupling between the dot and the 
leads, $G_{wc}$, with the lowest orders in the perturbation theory.  The 
expression for the capacitance far from the peak center is \cite{GlazmanMatveev,weak}: 
\begin{equation}
C=\frac{\partial Q}{\partial V_{g}}=aC_{gd}G_{wc} \!\! \left( \frac{h}{4 \pi ^{2}e^{2}} \right) \!\! \left( \! \frac{1}{0.5-N} + \frac{1}{0.5+N} \!\! \right)
\end{equation}
Near the peak center, the calculation for non-zero temperatures yields an 
expression with a Fermi-Dirac component and a correction that is linearly 
dependent on $G_{wc}$ \cite{weakpeak}. In the theory \cite{weak}, $a=1$. 

For strong tunneling, the theory \cite{strong} is based on perturbation in 
the reflection amplitude. Here, the capacitance lineshape is:
\begin{equation}
C(N) \! = \! bC_{gd}r^{2} \ln \!\! \left( \! \frac{1}{r^{2} \cos ^{2} \pi N} \! \right) \!\! \cos 2 \pi N \! \! + \! C_{0}
\label{eq:strong}
\end{equation}
The reflection coefficient, $r$, is related to the tunnel barrier conductance as: 
$G_{sc}=2(1-r^{2})e^{2}/h$.  $C_{0}$ is a constant, determined by 
normalizing the integral of $C(N)$ to one electron, $b=2.27$. The logarithmic
divergence is analogous to a similar behavior of magnetic susceptibility in the
two-channel Kondo problem \cite{Kondo}.  To account for a finite temperature, the 
singularity in (\ref{eq:strong}) is cut off by replacing $r^{2} \cos ^{2} \pi  N$
with $r^{2} \cos ^{2} \pi  N + \frac{k_{B}T}{U}$.  The corrected expression was used
for the fits.

We fit every measured capacitance peak with the above described expressions 
using the conductance as a parameter with least squares optimization. In Fig. 
3a and the lower plot of Fig. 3b, we show fits for weak tunneling with 
conductances of $G_{wc}=0.55$ and 1.08$e^{2}/h$. These peaks are in 
excellent agreement with our data measured with tunnel barrier conductances 
of 0.67 and 1.09$e^{2}/h$, respectively. The strong tunneling lineshapes are 
shown in the top plot of Fig. 3b and in Figs. 3c and d.  These figures show the 
strong tunneling calculations for conductances of $G_{sc}=1.08$, 1.41 and 
1.90$e^{2}/h$.  There is excellent agreement of these lineshapes with our data, 
obtained with conductances of 1.09, 1.50 and 1.81$e^{2}/h$. Both the weak 
and strong coupling theories fit well to the capacitance lineshape obtained for 
$G=$1.09$e^{2}/h$, shown in Fig. 3b. Fig. 3d shows a capacitance peak for a 
nearly transparent tunnel barrier conductance.  Here, the lineshape is almost 
indistinguishable from a sinusoid and it is difficult to discern any significant 
differences between any of the theoretical calculations.

In the case of weak coupling, we found that on average, $G_{wc}$ corresponds to 
the experimentally measured value if $a=4$. For strong coupling, we found that the 
coefficient $b=1$, to maintain the dependence of the capacitance lineshape on 
$G_{sc}$ in this limit. Similar discrepancies were observed elsewhere \cite{Duncan},
but their cause is not known at this time.

Figure 3e shows the dependence of the tunnel barrier conductance on the voltage 
of the lead defining the tunnel barrier. We also plot the conductance values 
obtained from theoretical fits in the weakly and strongly coupled regimes.  
These values have large fluctuations around the measured tunnel barrier 
conductance. These fluctuations are a mystery that remains to be solved.  They 
are seen consistently in all of our samples. Evidently, for a dot with a single 
point contact, the tunnel barrier conductance affecting the lineshape is different 
from the conductance through the dot which does not display comparable 
fluctuations. In gate voltage sweeps for a fixed tunnel barrier conductance, the 
values of $G_{wc}$ or $G_{sc}$ are correlated over a few adjacent peaks. A similar 
effect was observed in conductance measurements in dots in the quantum 
Hall regime \cite{condlarge}.  Theorists predict that such fluctuations can arise 
from quantum interference inside the dot and should therefore 
be highly sensitive to magnetic field. This is consistent with results from 
{\it conductance} experiments \cite{condsmall}. There are similar theoretical 
predictions for fluctuations in capacitance peaks \cite{Aleiner}. However, we 
observed no effect of magnetic field for magnetic fluxes through the dot as 
high as 30 flux quanta. 

We thank K. Matveev, L. Levitov, and L. Glazman for numerous helpful 
discussions. This work is supported by the Office of Naval Research, the 
National Science Foundation DMR, the David and Lucille 
Packard Foundation, and the Joint Services Electronics program.

\end{multicols}

\begin{references}
\bibitem{Weisskopf} V. F. Weisskopf and E. Wigner, Z. Physik {\bf 63}, 54 (1930).
\bibitem{overview} R. C. Ashoori, Nature {\bf 379}, 413 (1996); 
M. A. Kastner, Physics Today {\bf 46}, 24 (1993);
L. P. Kouwenhoven {\it et al.}, Science {\bf 278}, 1788 (1997).
\bibitem{Devoret} M. Devoret and H. Grabert, in {\it Single Charge Tunneling}, edited by H. Grabert and M. Devoret
NATO ASI, Ser. B, {\bf 294} (Plenum, New York, 1992).
\bibitem{GlazmanMatveev} L.I. Glazman and K.A. Matveev JETP {\bf 71}, 1031 (1990).
\bibitem{weak} K. A. Matveev, Sov. Phys. JETP {\bf 72}, 892 (1991).
\bibitem{strong} K. A. Matveev, Phys. Rev. B. {\bf 51}, 1743 (1995).
\bibitem{Foxman} E. B. Foxman {\it et al.}, Phys. Rev. B {\bf 47}, 10020 (1993).
\bibitem{Furusaki} A. Furusaki and K. A. Matveev, Phys. Rev. Lett. {\bf 75}, 709 (1995).
\bibitem{Goldhaber} D. Golhaber-Gordon {\it et al.}, Nature {\bf 391},156 (1998).
\bibitem{Molenkamp} L. W. Molenkamp, K. Flensberg, and M. Kemerink, Phys. Rev. Lett. {\bf 75}, 4282 (1995).
\bibitem{Livermore} C. Livermore {\it et al.}, Science {\bf 274}, 1332 (1996).
\bibitem{MatveevGlazman} K. A. Matveev, L. I. Glazman and H. U. Baranger, Phys. Rev. B {\bf 54}, 5637 (1996).
\bibitem{Likharev} K. K. Likharev, IEEE Trans. Magn. {\bf 23}, 1142 (1987).
\bibitem{Berman} D. Berman {\it et al.}, J. Vac. Sci. Technol. B {\bf 15}, 2844 (1997).
\bibitem{Fulton} T. A. Fulton and G. J. Dolan, Phys. Rev. Lett. {\bf 59}, 109 (1987).
\bibitem{Lafarge} P. Lafarge {\it et al.}, Z. Phys. B. {\bf 85}, 327 (1991).
\bibitem{weakpeak} K. A. Matveev, private communication, 1997.
\bibitem{Kondo} N. Andrei and C. Destri, Phys. Rev. Lett. {\bf 52}, 364 (1984).
\bibitem{Duncan} D.S. Duncan and R.M. Westervelt, private communication, 1998.
\bibitem{condlarge} P.L. McEuen {\it et al.}, Phys. Rev. Lett. {\bf 66}, 1926 (1991).
\bibitem{condsmall} R.A. Jalabert, A.D. Stone and Y. Alhassid, Phys. Rev. Lett. {\bf 68}, 3468 (1992);
A.M. Chang {\it et al.}, Phys. Rev. Lett. {\bf 76}, 1695 (1996);
J.A. Folk {\it et al.}, Phys. Rev. Lett. {\bf 76}, 1699 (1996).
\bibitem{Aleiner} I.L. Aleiner and L.I. Glazman, cond-mat/9710195 (to appear in Phys. Rev. B
{\bf 57}, 1998); A. Kaminsky, I.L. Aleiner, and L.I. Glazman, cond-mat/9802159.
\end{references}
\end{document}